# A very accurate hard sphere equation of state over the entire stable and metstable region

Hongqin Liu

CRCT, Ecole Polytech., University of Montreal

Box 6079, Station Downtown, Montreal, QC, Canada, H3C 3A7

Email: hongqin.liu@polymtl.ca

The hard sphere system plays a basic role in condensed matter physics and related fields, and equation of state (EoS) is the ultimate solution to its thermodynamic properties [1,2,3]. It has been a challenge to develop an analytical EoS for the entire stable and metstable region [4]. By virtue of potential energy landscape analysis combined with the Woodcock EoS [2,5], here we show that an EoS can be obtained to reproduce the compressibility of the entire region with high accuracy. Therefore, the pressure of all four amorphous states of matter, gas, liquid, supercooled liquid and glass can be represented with a single EoS. Examples are given to show that highly accurate EoS is necessary for the predictions of thermodynamic or structural properties. By using conventional approaches, such as van der Waals theory [6] or perturbation theory [1,7], the EoS can be extended to real systems, including supercooled liquids and glasses.

The hard sphere (HS) system is a model system with pure repulsive interaction. This system finds diverse applications in stable gases and liquids, colloidal suspensions, metstable supercooled liquids and glasses, and granular matters etc. [1-3]. The study on HS EoS has been a very active subject over decades. A recent review covers a detailed comparison between about two dozens of EoSs [3]. Another dozen might be found in the literature in addition to those reported in the review. References [4,8,9,10] are among the most recent works. A "well-behaved" HS EoS might be interpreted as meeting several requirements. (1) It should be very accurate, necessary for many applications. For instance, in the perturbation theory for real fluids, the HS system is used as a reference and additional uncertainty will be created by the theory. Deviations in the HS EoS will be carried forward into the real system and cause confusion, as shown later. (2) A HS EoS should reproduce as many as possible exact or accurate virial coefficients. This is because the virial equation of state is the only equation with sound theoretical background, and can produce the most accurate compressibility data and other thermodynamic properties in the stable fluid region, given sufficient virial expansion terms. (3) The equation should be fairly simple for mathematical manipulations since other thermodynamic properties will be derived from it by some calculus. For the stable fluid region, the widely used Carnahan and Starling (CS) EoS [11] is a good example of being both simple and accurate:

$$Z = \frac{1+\eta+\eta^2-\eta^3}{(1-\eta)^3} \qquad (1)$$

where $Z(=PV/kNT)$ is the compressibility, and $P$, the pressure, $V$, the total volume, $T$, the temperature, $N$, the total number of particles, $k$, the Boltzmann constant; $\eta$, the packing fraction, defined as $\eta = \pi\rho^*/6 = \pi\rho\sigma^3/6$, $\rho(=N/V)$, the number density and σ, the hard sphere diameter. Some very accurate EoSs are available for the stable region and a particular example is the equation proposed by Kolafa et al. (here Eq.(10) in the article [10] will be tested for comparison), which is a truncated virial expansion with adjustable high order terms. For the HS glass Speedy used a simple EoS [12,13]:

$$Z = \frac{a}{1-\alpha\eta} \qquad (2)$$

where *a* and α are constants. Despite its simplicity, Eq.(2) is the most accurate EoS available now in the metstable region. Unfortunately, it is not accurate enough for some applications. It is easy to show that Eq.(2) leads to a constant heat capacity, while an accurate calculation shows dramatic changes in the metstable region [14]. Most importantly, as shown later, using two EoSs for the stable and metstable region, respectively, is not only inconvenient, but also sometimes misleading.

Many efforts have been made aiming at a unified EoS over the entire density range or to deep supercooled liquid densities [1,2,3,4,10,15,16,17]. The Woodcock EoS [2,5] is of particular interest to the present study. It is composed of a truncated virial expansion part and Eq.(2), which diverges at the random close packing (RCP) density (recently defined as the maximally random jammed, MRJ, density [18]). The Woodcock EoS could not work well for both stable and metstable regions simultaneously. In fact all EoSs for the whole density range suffer the same shortcoming. The Le Fevre EoS [15] is still one of the best EoSs for the entire region, but its accuracy is not satisfactory as shown later. Obviously, there is something missing in these "classic" EoSs.

Here, we revisit this issue by virtue of the potential energy landscape (PEL) formulism [19,20,21,22]. The total pressure is expressed as the sum of high temperature contribution ($Z_{HT}$) and low temperature contribution ($Z_{LT}$), and the later is decomposed into two parts: the inherent structure contribution ($Z_{IS}$) and the vibrational contribution ($Z_{vib}$): $P = P_{HT} + P_{LT} = P_{HT} + P_{IS} + P_{vib}$ (see Method for details). The final EoS reads:

$$Z = 1 + \sum_{i=1}^{12} a_{i+1}\eta^i + \frac{c_0 \eta}{1-\alpha\eta} + c_1\eta^{40} + c_2\eta^{42} c_3\eta^{44} \qquad (3)$$

As seen from the equation, the Woodcock type EoS [5] covered $P_{HT}$ and $P_{IS}$ contributions, but missed $P_{vib}$. The constants in Eq.(4) were determined as follows. The coefficients $a_i$ ($i = 1,2,\cdots 12$) were calculated such that all the values or the virial coefficients up to the 12$^{th}$ reported in reference [23] can be reproduced, and the 13$^{th}$ is re-estimated in this work. In another recent publication (Labik et al., 2005) [24], the virial coefficients up to the 10$^{th}$ are also reported, and some values are slightly different from those reported in reference [23]. No considerable differences in the final results were found when the virial coefficients from reference [24] were adopted. The values of constants $c_i, (i = 0,1,2,3)$, α and the powers (40, 42, 44) were adjusted parameters for the best fit of the compressibility data over the entire density range. Two objective functions, composed of the average absolute deviations or standard deviations [10], respectively, were adopted and no considerable differences in the results were found. Finally, the relation between virial coefficient, $B_i$ and coefficient $a_i$ is simply given by $B_i = a_i + c_0\alpha^{(i-2)}$ ($i = 2,\cdots 12$).

By the way, the truncated virial term of Eq.(3) can be reformulated into the following compacted equation:

$$Z_v = 1 + \frac{3.68584\eta}{1 - 2.5848\eta + 1.9499\eta^2 - 0.172284\eta^3 - 0.16012\eta^4} \qquad (4a)$$

Then Eq.(3) can be rewritten as:

$$Z = Z_v + \frac{c_0 \eta}{1 - \alpha\eta} + c_1\eta^{40} + c_2\eta^{42} + c_3\eta^{44} \qquad (4b)$$

It is found that both Eq.(3) and Eq.(4) give almost exactly the same results for compressibility and chemical potential, while the accuracies of the virial coefficients derived from Eq.(4), noted as $b_i$, are sacrificed somewhat as $i>9$, but yet acceptable. The values of the constants in Eq.(3) and Eq.(4) are given in Table 1.

The most important parameter for the metstable region is $\alpha$: its inverse gives the maximally random jammed packing: $\eta_J = 1/\alpha$. As discussed below and elsewhere [14], the behavior of a glass is closely related to this quantity. The result from the present work is $\eta_J = 0.635584$, ($\rho_J^* = 1.21388$), which is very close to the values obtained with other means. The value observed by Scott and Kilgour [25] using a physical experiment is 0.6366, and same value was also obtained by Finney [26]. A recent computer simulation result is 0.64 [18].

The compressibility data are crucial for the parameter regression. Fortunately, a large number of MC and MD simulation data points are available in the literature. The data have been carefully evaluated and, for the stable region, compared with the predictions of the truncated virial equation [23]. For the stable region ($\rho^*$ = 0 to 0.95), 64 data points were used: Erpenbeck & Wood (1984, 10 points) [27], Kolafa et al. (2004, 22 points) [10], , Wu & Sadus (2005, 32 points) [28]. For the metstabe region ($\rho^*$=0.95 to 1.21), 81 data points were adopted: Woodcock (1981, 15 points) [2], Rintoul & Torquato (1996, 12 points, one point not used at $\rho^* > 1.21$) [16], Speedy (1997, 17 points) [29], Kolafa et al. (2004, 9 points), Wu & Sadus (2005, 32 points, one point not used at $\rho^* > 1.21$) [28], Kolafa (2006, 8 points) [4]. The rejection of two points at $\rho^* > 1.21$ is due to large uncertainty very close to the maximally jammed density ($\rho_J^* = 1.21388$). Since the values of $\eta_J$ is "history" dependent, the pressures at the densities very close to the point should be also vary. All data points along with the calculation results are listed in Supplementary Materials.

Table 2 summarizes the results. As shown in the table, Eq.(3) or Eq.(4) is as accurate as the Kolafa et al. EoS [10] in the stable fluid region. This is not surprising since the new EoS reproduce accurate virial coefficients up to the 12$^{th}$. The CS EoS is less accurate due to the fact that it was derived from the approximate virial coefficients [11]. In the metstable region, the present EoS is significantly better than previous EoSs. The uncertainty of the new EoS is well within the discrepancies between the different data sources.

Figure 1a and 1b depict the comparison of calculated compressibility with simulation data. The results lead to the conclusion that the pressure of all four amorphous states of matter, *gas, liquid, supercooled liquid and glass (corresponding to a specific $\eta_J$* *value) is a continuous function of density and can be very well represented by a single equation of state*.

Figure 2 shows the relative deviations (%) for the stable and metstable regions. For the stable region, the maximum deviation is about 0.48% from Eq.(3) or from the Kolafa et al. EoS. For the metstable region, deviations of Eq.(3) is well distributed around zero, and the maximum deviation is 6.7%. The other two EoSs show systematic errors.

We emphasize that eq.(3) or (4) is not unique for the HS compressibility (pressure). In the PEL framework, the pressure at very low temperature is dependent on $\eta_J$ [12,30]: $P_{LT} = P_{LT}(\eta; c_i\{\eta_J\}, \eta_J)$, where $c_i\{\eta_J\}$ ($i$=0,1,…) denotes that the parameters in Eq.(3) are dependent on $\eta_J$. The value of packing fraction at the MRJ state, $\eta_J$, depends on the protocol employed to produce the random packing as well as other system characteristics [18,30]. This "history" dependent feature is consistent with the non-equilibrium properties of glasses. Different glasses have been reported in the HS system [12,30]. Then a natural question is: is it meaningful to propose an EoS for the whole metstable region? The answer is positive. Firstly, as in the present case, for a given $\eta_J$ corresponding to a specific glass, an analytical EoS proves the continuity of the pressure of all four amorphous states of matter. Secondly, as different glasses are concerned, it is found that the derived properties of supercooled liquids and glasses, such as heat capacity, are repeatable up to a packing fraction, $\eta \approx 0.61$. Therefore, the new EoS can be used as a very useful tool not only for providing very accurate thermodynamic properties for the stable region, but also for predicting or analyzing the properties of supercooled liquids and glasses [14].

Figure 3 illustrates the relationship between the stable fluid, the crystal solid and the glasses by using the inverse of the radial distribution function at contact calculated with different EoSs, $G = (Z-1)/4\eta$. Several important facts can be observed from this figure. (1) The CS EoS is not applicable as packing fraction is greater than 0.55 and the Speedy EoS shows considerable deviation between 0.56 and 0.61. Therefore, due to glass transition of the HS system appears at $\eta \approx 0.58$ [14], it is not appropriate to use the intersection of $1/G$ from the CS EoS with that from the Speedy EoS as a signature of glass transition, as did by reference [31]. (2) Although the CS EoS is not applicable for a non-equilibrium glass transition, it has been used to produce "equilibrium" glass transition: the dashed line extrapolates the CS EoS to 0.68, the equilibrium maximally random packing [30]. In comparison, the present EoS gives one of the non-equilibrium glasses at the MRJ state at $\eta \approx 0.6356$. By the way, the close packing of the crystal solid is at $\eta \approx 0.74$. (3) It is possible to get some other EoSs, which can satisfactorily reproduce all simulation data (possibly with few exceptions as very close to $\eta_J$) while end at different value of $\eta_J$. For example, one may reasonably extrapolate the data point along the dotted line (the Speedy EoS) to $\eta_J = 0.646$. Such a difference may have no impact on the properties of stable region and supercooled liquids, but it indeed has significant impact on the properties of glasses as the glass density varies within a small range.

For a further test of the EoS, as suggested by Mulero et al. [32], chemical potential has been calculated with the equation: $\mu = Z - 1 + \int_0^\eta (Z-1)d\eta/\eta$. The calculated results are compared with the simulation data (in stable fluid range) [33,34,35] and the results are also given in Table 2. Again, Eq.(3) shows a high accuracy.

Why is such a high accuracy necessary for a HS EoS? It is demonstrated elsewhere that a high accuracy is crucial for glass transition study [14]. Here we show two examples in the stable fluid region, where the HS EoS is applied in predicting thermodynamic and fluid structure properties, respectively.

As mentioned previously, an important application of a HS EoS is in the perturbation theory [1], and a good example is the Weeks-Chandler-Andersen (WCA) theory for the Lennard-Jones (LJ) fluid [36,37]. In this theory, the LJ potential is decomposed into two parts: a repulsive part (RLJ) and an attractive part. The repulsive part uses the HS system as reference, in which the particle diameter, σ, is replaced by a so called effective diameter $\sigma_{eff}(T, \rho)$, then the pressure of the RLJ fluid is calculated by a HS EoS with the packing fraction calculated with $\frac{1}{6}\pi\rho\sigma_{eff}^3$. There are different ways to determine $\sigma_{eff}(T, \rho)$ according to different perturbation theories. Here we consider three cases: (1) the Verlet & Weis (VW, $\sigma_{eff}$ is only temperature-dependent) [38]; (2) the WCA theory [37] ($\sigma_{eff}$ is both temperature and density dependent); (3) the Lado theory [39] ($\sigma_{eff}$ is both temperature and density dependent). The analytical expressions for $\sigma_{eff}(T, \rho)$ from each theory can be found in references [7,38]. We adopted the simulation data for the pressure of RLJ reported by Mulero et al. [3]. Three EoSs, the CS EoS, Eq.(1), the new EoS, Eq.(10), and the Le Fevre EoS [15], were employed. The results are listed in Table 3.

What can we learn from these results? An immediate observation from Table 3 tells that all three EoSs show the Lado theory giving the best results. Indeed, this theory [39] overcame a drawback in the WCA theory [37], and, not surprisingly, has a better performance.

If merely by looking at the AADs from different EoSs, one might suggest that the Le Fevre EoS gives the best results, and according to this EoS, the Lado theory is unusually accurate, AAD=0.58%. This is of course misleading. We already know that for the HS fluid in the stable region (as in the present case), the Le Fevre EoS is not as good as Eq.(1) or Eq.(3). How can it give "better" results for the RLJ fluid which is actually treated as a HS system using exactly the same EoS? The reason for this unusual accuracy is that the error of the Le Fevre EoS coincidently cancels the error of the perturbation theory. In fact, it is found that in the stable region, the Le Fevre EoS underestimates the HS pressure (about 0.5%) while the Lado theory overestimates the RLJ pressure. The result fom Eq.(3) shows that for the Lado theory, the overestimated deviation (AAD) is about 1.3%.

The above case demonstrates that an accurate HS EoS is necessary for thermodynamic property calculations. The same is true for the liquid structure prediction.

A HS EoS is involved in the calculation of the direct correlation function (DCF) and structure factor. Here the theory of Baus & Colot [40] is adopted. An EoS is required to calculate the DCF at the end point: $c(x=0;\eta) = c(q=0;\eta) = -\partial(\eta Z)/\partial\eta$, where $x$ and $q$ are the variables in the real space and the Fourier space, respectively. This approximation omits a positive "tail" of the CDF [41]. It is found that at high densities, such an approximation is not satisfactory. Consequently, a slight modification is introduced here for an accurate prediction, and we use the following equation to calculate the DCF [40]:

$$c(x;\eta) = -D\,\Theta(1-x)\left[1 + \frac{1}{2}\eta x^3 - \frac{x}{6\eta}\left(2\eta^2 + 8\eta - 1 + \frac{1}{D}\right)\right] \quad (5)$$

where $D = 1.1[\partial(\eta Z)/\partial\eta]$ and $\Theta(1-x)$ is the Heaviside step function. The factor "1.1" is an empirical correction to the original approximation. The Fourier transform of Eq.(5) gives $c(q;\eta) = \frac{4\pi}{q}\int x c(x;\eta)\sin(qx)dx$, and finally the structure factor is calculated by

$$S(q;\eta) = \frac{1}{1 - c(q;\eta)}. \quad (6)$$

Figure 4 shows the calculated structure factors by using the CS EoS, Eq.(1) and the new EoS, Eq.(3). The simulation data are from reference [42]. As seen from the figure, at low and intermediate densities, both EoSs give almost the same results. But at high densities, such as at $\rho^* = 1.04$, where the CS EoS still gives reasonable value for Z, while underestimates the derivative D, the difference in structure factors form the two EoSs becomes considerable. Figure 4 also shows that at the high density (1.04) the liquid structure is experiencing a dramatic change: a precursor of a glass transition [14].

**Method**

In the PEL picture, the system composed of N particles is distributed on a 3N+1 dimension potential energy surface with huge number of local minima, which is known as inherent structure (IS), and this allows a decomposition of the partition function into an IS part (inter-basin), connected to the zero temperature landscape corresponding to the configurations of the system at temperature T, and a part (intra basin), connected to the thermal excitation of the configuration in a single minimum (vibration) [19]. Stillinger and Weber first formulated the partition function within the PEL framework [20]. Only in recent years, efforts have been made on deriving an EoS based on the PEL approach [21,22]. In the PEL formulism, the partition function of the system, $Q$, dealing with 3N+1 dimension potential energy surface, is simply expressed as the integration of one single variable, basin depth, $\phi$ [20,21]:

$$Q = e^{-\beta A} = C \int_{\phi_{\min}(\rho)}^{\phi_{\max}(\rho)} \exp[N(\sigma(\phi,\rho) - \beta\phi - \beta a_{vib}(\beta,\phi,\rho))]d\phi \qquad (7)$$

where A is the free energy, $\beta = 1/T$, C is a constant, $\sigma(\phi,\rho)$, the basin enumeration function, $a_{vib}(\beta,\phi,\rho)$, the basin vibrational free energy when the system is confined to an average basin of depth $\phi$, and finally, the basin enumeration function is defined such that $C\exp[N\sigma(\phi,\rho)]d\phi$ gives the number of inherent structures with potential energy per particle $\phi \pm \frac{1}{2}d\phi$.

At very low temperature (high density for HS system), the system is in metstable state (supercool liquids or ideal glasses). Then in a large system limit, the integral in Eq.(7) will be dominated by a maximum value of exponential term at some $\phi = \phi^*$. This means that, at the given density (and temperature), the system will sample configurations whose overwhelming majority have energy $\phi^*$ [22]. Then the system free energy can be derived from Eq.(7):

$$A/N \approx [\phi^* - kT\sigma(\phi^*, \rho)]_{IS} + a_{vib}(\beta, \phi^*, \rho) \qquad (8)$$

Accordingly, the pressure for the metstable system can be written as

$$P_{LT} = P_{IS} + P_{vib} \qquad (9)$$

where the subscript, $LT$, refers to low temperature since both IS and vibrational contributions become significant only in supercooled liquids and glasses [19]. Apparently, for an analytical EoS from Eq.(8) and Eq.(9), one needs to know the dependences of $\sigma(\phi^*, \rho)$ and $a_{vib}(\beta, \phi^*, \rho)$ on the IS structure and the issue is discussed in references [19,20]. Although the density-dependences of the structure parameters are generally unknown, the qualitative discussions within the PEL framework already provide very useful information for us to proceed.

Since our goal is to derive an equation covering the whole density (temperature) range, we express the system pressure as $P = P_{HT} + P_{LT}$, where "$HT$" refers to high temperature, which is closely related to the configurational contribution. The equation should be read as such: at high temperature, $P_{HT}$ dominates, while at low temperature, $P_{LT}$ dominates. For the high temperature contribution, the virial expansion is a handy candidate, then the total pressure can be written as

$$P = P_{virial} + P_{IS} + P_{vib} \qquad (10)$$

where $P_{virial} = 1 + \sum_{i=2}^{m} B_i \eta^{i-1}$, $B_i$, the virial coefficient. In the PEL framework, at zero temperature (MRJ state), the whole system is trapped within its inherent structure (basins), which is reflected by the divergence of $P_{IS}$ at the MRJ density. The simplest function for this part is naturally Eq.(2), which is strict in the one dimension case and has been proved quite satisfactory for the metstable HS system [5,13]. Some theoretical justification has also been suggested [30]. Therefore, the following equation is adopted:

$$P_{IS} = a\left[\frac{1}{1-\alpha\eta} - \sum_{i=0}^{m}(\alpha\eta)^i\right] \quad (11)$$

where $a$ and $\alpha$ are constants. The second part of Eq.(11) is for subtracting the high temperature contribution which has already been covered in the truncated virial expansion term.

Now we need an expression for $P_{vib}$. From the above PEL analysis, we can see that this component has two features: (1) at high temperature its contribution is negligible; (2) at low temperature, its contribution is considerable, but becomes less important as $T \to 0$ since it does not diverge. Considering these requirements, a simple choice is a high order polynomial function: $P_{vib} = \sum_{i>n_0}^{l} c_i \eta^i$. The selection of constants $l$ and $n_0$ is somewhat arbitrary, and we leave them to be determined by fitting measured compressibility data. The final EoS is given by Eq.(3).

**Acknowledgement**

I thank Dr. A. Mulero, Dr. J. Kolafa, and Dr. Robin J Speedy for valuable discussions and suggestions during the preparation of the manuscript. I also thank Dr. L. V. Woodcock, and Dr. S. Torquato for answering my questions regarding the related subjects and providing their articles. I am grateful to Dr. A. Pelton and Dr. S. Decterov (CRCT, Ecole Polytech.) for their support.

**Competing Interests**

The author declares no financial interests.

**Figure captions**

**Figure 1**. Comparison of the calculated compressibility with the simulation data.

**a**. The stable fluid region.

The numbers in the parentheses are the reference numbers. The Kolafa et al. EoS is their Eq.(10) in the paper [10]. The CS EoS gives accurate compressibility up to around $\rho^*=1.05$. The Kolafa et al. EoS can be used up to 1.12.

**b**. The entire density range and the metstable region.

The symbols are the same as in Figure 1a. The inner panel shows the results for the density range from 0.9 to 1.14.

**Figure** 3. Plot of the inverse of the radial distribution function at contact. The EoS for crystal solid is given in reference [2]. The extrapolation of CS EoS to packing fraction 0.68 was adopted from reference [30], as an "equilibrium" glass transition.

**Figure 4**. The structure factor. The numbers in the figure refer to the reduced density, $\rho^*$. Solid lines are from Eq.(3), and dashed lines from the CS EoS [11]; the points are MD simulation data from Alley & Alder (1983) [42].

**Table 1** values of constants

| $i$ | 1 | 2 | 3 | 4 | 5 | 6 | 7 |
|---|---|---|---|---|---|---|---|
| $a_i$ | 1 | 3.68584 | 9.50571 | 17.58708 | 27.00093 | 37.89002 | 50.3155 |
| $B_i$ | 1 | 4 | 10 | 18.36477 | 28.2245 | 39.81515 | 53.3442 |
| $b_i$ | 1 | 4 | 10.0214 | 18.2165 | 28.3573 | 40.2881 | 53.8112 |

| $i$ | 8 | 9 | 10 | 11 | 12 | 13 |
|---|---|---|---|---|---|---|
| $a_i$ | 63.7720 | 78.3149 | 93.97817 | 109.3655 | 123.4699 | 172.272 |
| $B_i$ | 68.53755 | 85.81284 | 105.7751 | 127.9263 | 152.6727 | 218.31* |
| $b_i$ | 68.6907 | 84.6649 | 101.502 | 119.107 | 137.708 | 158.167 |

| $c_0$ | $c_1$ | $c_2$ | $c_3$ | $\alpha$ |
|---|---|---|---|---|
| 0.31416 | $4.1637\times 10^{10}$ | $-2.3452\times 10^{11}$ | $3.6684\times 10^{11}$ | 1.573357 |

Notes: $\eta_J = 1/\alpha = 0.635584$. $B_i$s are the virial coefficients from reference [23] and can be reproduced by $a_i$s. $b_i$s are the virial coefficients from the expansion of Eq.(4b), * fitted value.

**Table 2**. Summary of results

|  | Compressibility | | Chemical Potential |
|---|---|---|---|
| EoS | stable region | metstable region | |
|  | (64 points) | (81 points) | (35 points) |
| Eq.(3) | 0.0837 | 0.97 | 0.64 |
| Eq.(4) | 0.0841 | 0.98 | 0.64 |
| Kolafa EoS | 0.0834 |  | 0.64 |
| CS EoS | 0.196 |  | 0.71 |
| Le Fevre EoS | 0.592 | 8.36 | 1.07 |
| Speedy EoS |  | 4.70 |  |

Notes : the numbers in the table are AADs (%): $AAD = 100 N_P^{-1} \sum_{i=1}^{N_P} \left| \left( Z_i^{cal} - Z_i^{sim} \right) / Z_i^{sim} \right|$, where $N_P$ is the total number of data points. For the Speedy EoS, Eq.(2), the two parameters were regressed by the data in the metstable region: $a = 2.8$, $\alpha^{-1} = 0.64626$.

**Table 3** Repulsive LJ pressure prediction *

($T^* = 0.7 \sim 2.6$, $\rho^* = 0.25 \sim 0.844$, number of data points: 33) [3]

|  | Le Fevre EoS | CS EoS | New EoS |
|---|---|---|---|
| VW $\sigma_{eff}$ | 4.73 | 5.39 | 5.61 |
| WCA $\sigma_{eff}$ | 2.45 | 3.01 | 3.22 |
| Lado $\sigma_{eff}$ | 0.58 | 1.07 | 1.28 |

* the numbers in the table are the average absolute deviation (AAD, in percentage).

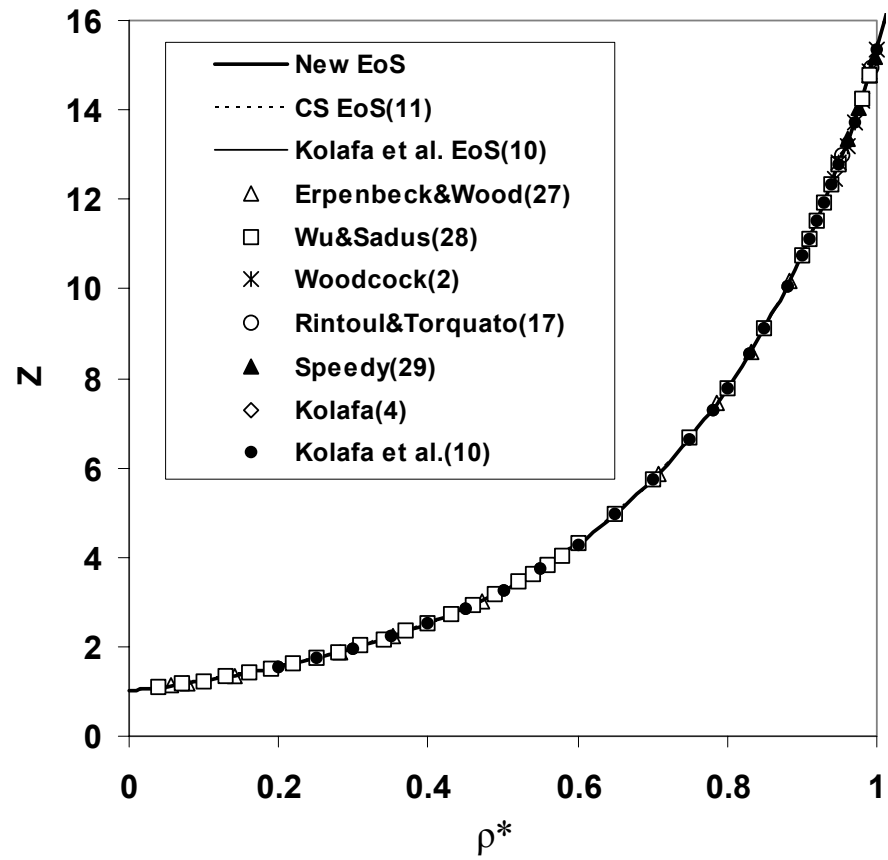

Figure 1a

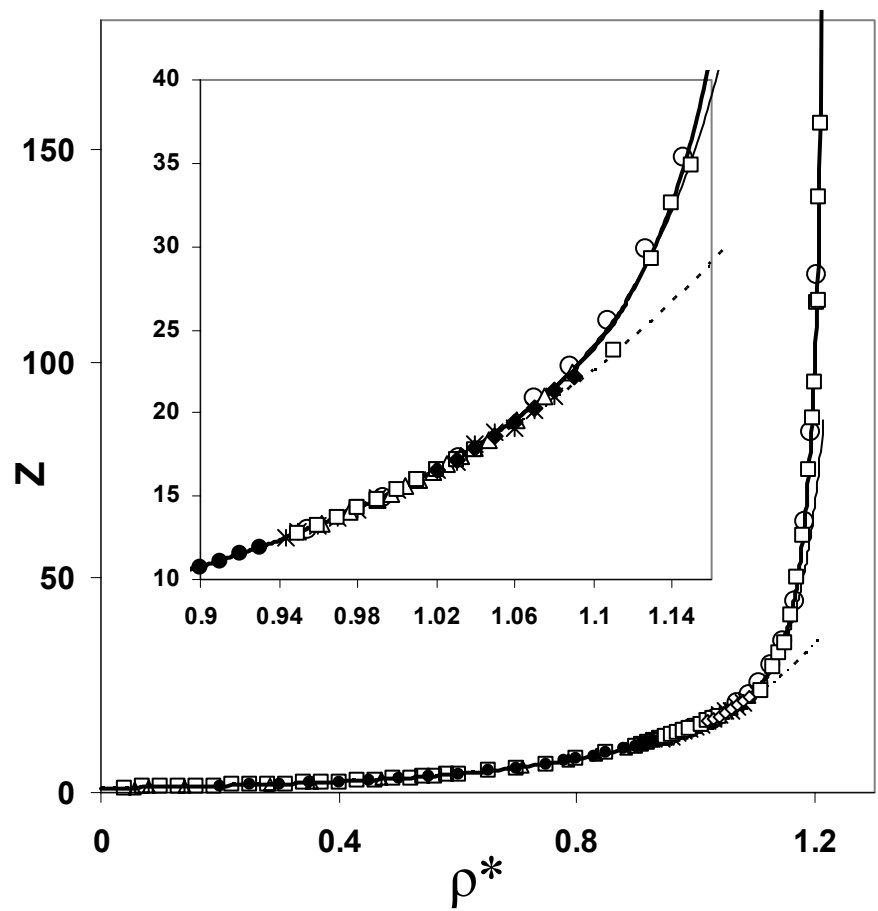

Figure 1b

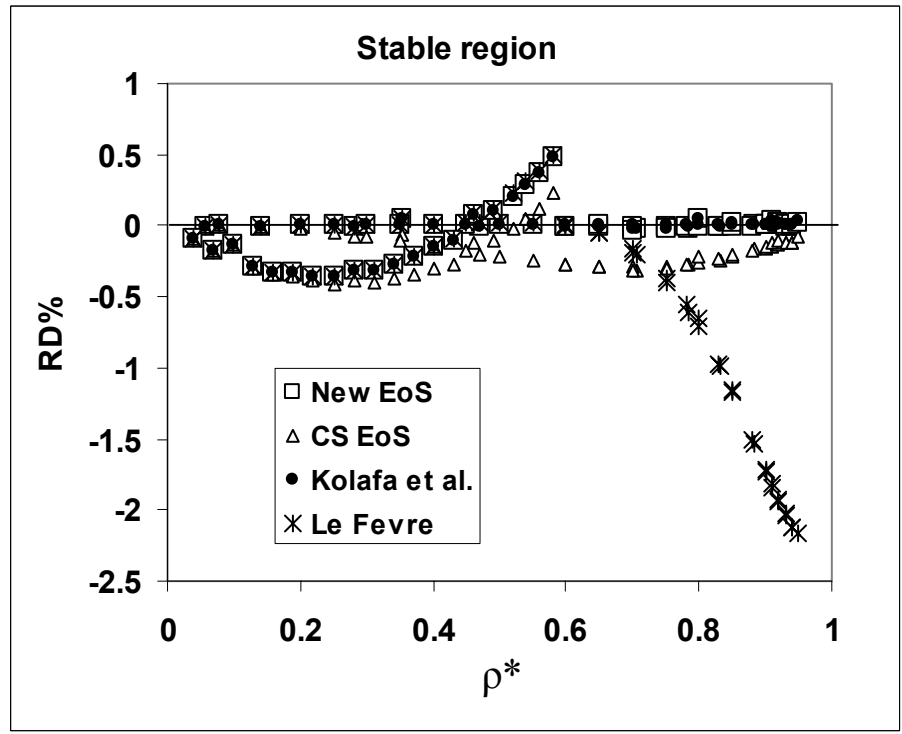

Figure 2a

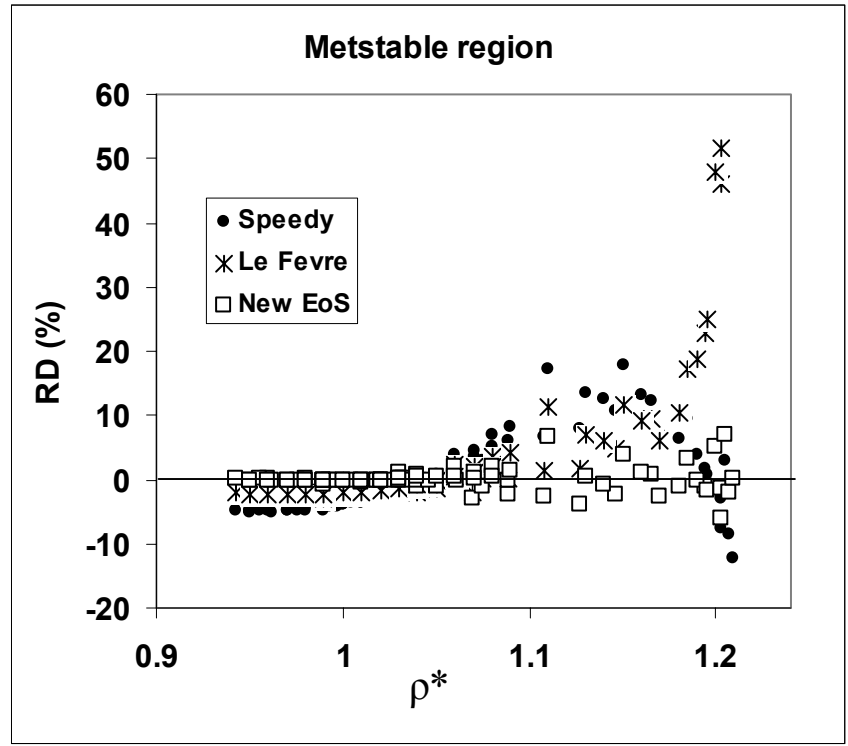

Figure 2b

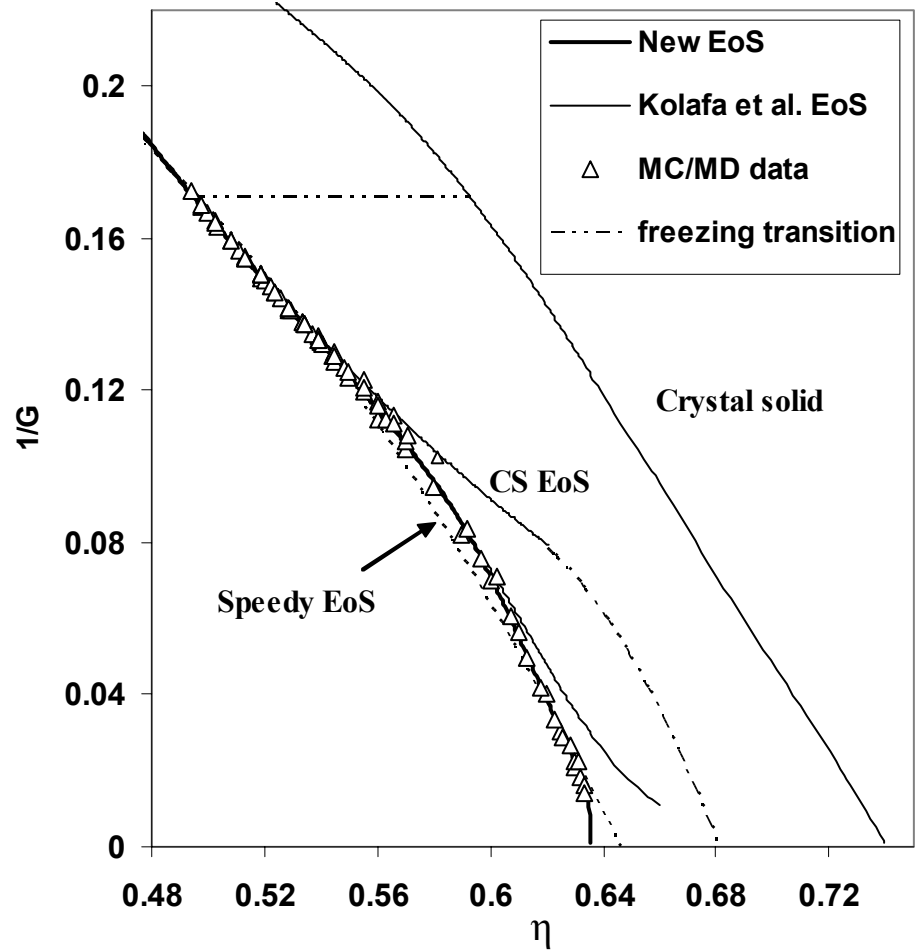

Figure 3

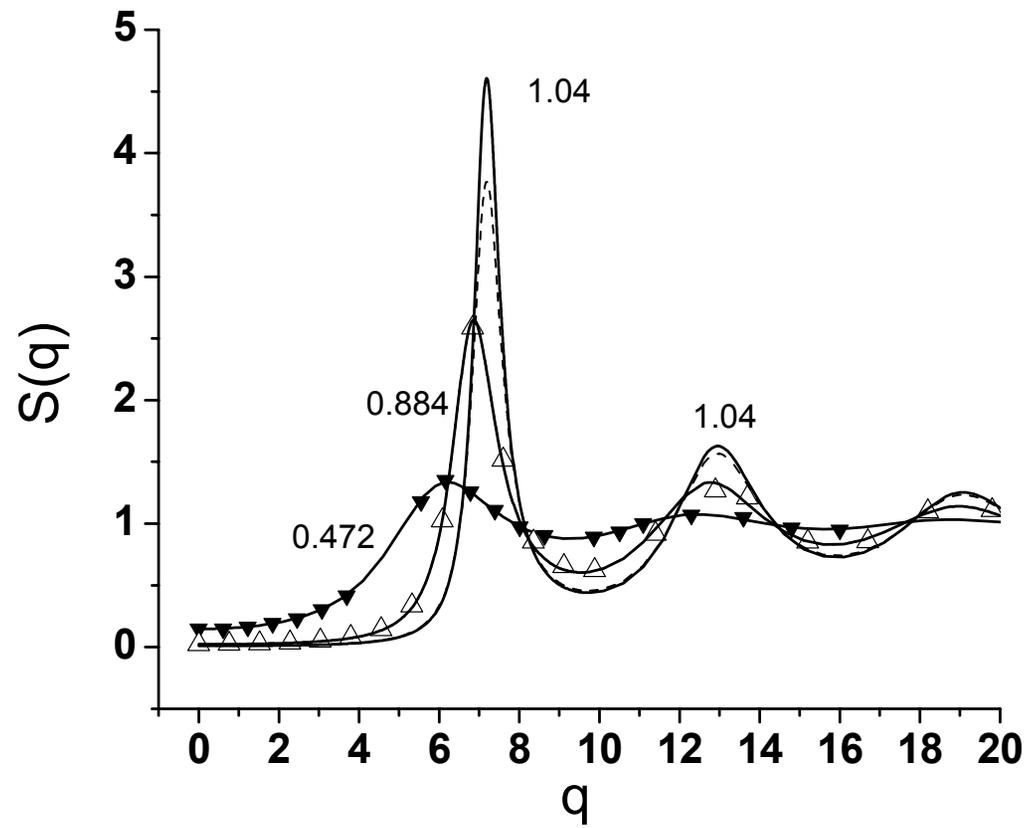

Figure 4